\documentclass[prl,twocolumn,twoside,floatfix,showpacs]{revtex4}
\usepackage{graphicx}
\usepackage{graphicx,amsfonts}
\usepackage{epsfig,amsmath}
\usepackage{verbatim,pslatex}

\begin{document}

\newcommand{\be}{\begin{equation}}
\newcommand{\ee}{\end{equation}}

\title{The quantum phase transition in the sub-ohmic spin-boson model:\\
Quantum Monte-Carlo study with a continuous imaginary time
cluster algorithm}

\author{Andr\'e Winter}
\affiliation{Theoretische Physik, Universit\"at des Saarlandes,
66041 Saarbr\"ucken, Germany}
\author{Heiko Rieger}
\affiliation{Theoretische Physik, Universit\"at des Saarlandes,
66041 Saarbr\"ucken, Germany}

\author{Matthias Vojta}
\affiliation{
Institut f\"ur Theoretische Physik, Universit\"at zu K\"oln, Z\"ulpicher Stra{\ss}e 77,
50937 K\"oln, Germany}
\author{Ralf Bulla}
\affiliation{
Institut f\"ur Theoretische Physik, Universit\"at zu K\"oln, Z\"ulpicher Stra{\ss}e 77,
50937 K\"oln, Germany}

\date{\today}

\begin{abstract}
A continuous time cluster algorithm for two-level systems coupled to a
dissipative bosonic bath is presented and applied to the sub-ohmic
spin-Boson model. When the power $s$ of the spectral function
$J(\omega)\propto \omega^s$ is smaller than $1/2$, the critical
exponents are found to be classical, mean-field like.
Potential sources for the discrepancy with
recent renormalization group predictions are traced back
to the effect of a dangerously irrelevant variable.
\end{abstract}
\maketitle


Quantum-mechanical systems embedded into a dissipative
environment play an important role in many areas of physics
\cite{legget-etal,weiss}. Among the numerous applications of models
that couple a small quantum-mechanical system to a bosonic bath are
noisy quantum dots \cite{quantum-dots}, decoherence of qubits in
quantum computations \cite{qbits}, and charge transfer in
donor-acceptors systems \cite{transfer}. A major research field are
quantum impurity models (i.e.\ a quantum spin embedded in a crystal
lattice, for a review see \cite{bulla-etal}), where in particular
quantum critical points occurring for
instance in the Bose-Fermi Kondo model have been studied intensively
\cite{kevin,kevin2,si,si2}.

The paradigmatic model of a two-state system
coupled to an infinite number of bosonic degrees of freedom is the
spin-boson model \cite{legget-etal,weiss}. As a function of the
strength of the coupling to its bath it displays a quantum phase
transition (QPT) at zero temperature between a delocalized phase,
which allows quantum mechanical tunneling between the two
states, and a localized phase, in which the system ceases to tunnel
in the low-energy limit and behaves essentially classically.

While the phase transition is understood in the case of
ohmic dissipation ($s=1$), the sub-ohmic situation ($s<1$) has been
investigated in detail only recently. On general grounds, one expects
the phase transition to fall into the same universality class as that
of the classical Ising spin chain with long-range interactions
\cite{suzuki}. Indeed, a continuous QPT has been found
in the spin-boson model for all values of $0<s<1$ \cite{BTV}, using a
generalization of Wilson's numerical renormalization group (NRG)
technique \cite{bulla-etal}. However, on the basis of these NRG
calculations, it was suggested that the quantum-to-classical mapping
fails for $s<1/2$ \cite{vojta}: There, the Ising chain displays a
mean-field transition, whereas the critical exponents extracted from
NRG were non-mean-field-like and obeyed hyperscaling.  Subsequent NRG
calculations for the spin-boson \cite{karyn} and Ising-symmetric
Bose-Fermi Kondo model \cite{kevin} confirmed this claim.  Such a
breakdown of quantum-to-classical mapping has consequences not only
for quantum-dissipative systems, but also for Kondo lattice models
studied within extended dynamical mean-field theory, where
non-mean-field critical behavior is at the heart of so-called local
quantum criticality \cite{si}.

The purpose of this letter is two-fold: 1) We present a novel and
accurate quantum Monte-Carlo (QMC) method to study the low temperature
properties of the sub-ohmic spin-boson model, and 2) we determine its
critical exponents at the quantum phase transition using this method
together with finite temperature scaling and re-confirm the
correctness of the quantum-classical mapping for the sub-ohmic bath
with $s<1/2$.

The spin-boson Hamiltonian is defined as
\be
H = \Delta\frac{\hat{\sigma}^x}{2}
+ \frac{\hat{\sigma}^z}{2}\,\sum_i \lambda_i({\bf a}_i+{\bf a}_i^+)
+ \sum_i\omega_i\,{\bf a}_i^+{\bf a}_i
\label{SB}
\ee
where $\sigma^{x,z}$ are Pauli spin-1/2 operators, ${\bf a}_i^+$,
${\bf a}_i$ are bosonic creation and annihilation operators, $\Delta$
the tunnel matrix element, and $\omega_i$ the oscillator frequencies
of the bosonic degrees of freedom.  The coupling between the the spin
$\sigma$ and the bath via the $\lambda_i$ is determined by the
spectral function for the bath:
\be
J(\omega)=\pi\sum_i \lambda_i^2\delta(\omega-\omega_i)
=2\pi\alpha\cdot\omega_c^{1-s}\omega^s
\label{spectral}
\ee
for $0<\omega<\omega_c$ and $J(\omega)=0$ otherwise.
$\alpha$ represents the coupling strength to the dissipative bath
and $\omega_c$ is a cut-off frequency. The parameter $s$ specifies the
low-frequency behavior of the spectral function: $s=1$ represents an
ohmic bath, and $s<1$ a sub-ohmic bath. A system described by
(\ref{SB}) and (\ref{spectral}) displays for $s\le1$ a quantum phase
transition (at zero temperature) at a critical coupling strength
$\alpha_c$. In the following we determine the critical exponents and
herewith the universality class of this transition with the help
of a continuous time cluster algorithm that samples stochastically
the imaginary time path integral for the partition function of
the model (\ref{SB}).

\newcommand{\ooo}{\underline{\hat{a}}}
\newcommand{\aaa}{\underline{{\omega}}}

Consider a Hamiltonian for an Ising spin in a transverse field
of the form
\be
{\it H}
=\Gamma \hat{\sigma}^x + {\it G}(\hat{\sigma}^z,\ooo,\aaa)\;,
\ee
where
$\Gamma$ is the transverse field strength, $\Gamma=\Delta/2$ in
(\ref{SB}), $\ooo$ and $\aaa$ a set of Hermitian operators and
parameters, respectively, like the Bose operators and coupling
constants and frequencies in the spin-boson model. ${\it G}$ is a
function of the $\hat{\sigma}^z$ and ($\ooo$, $\aaa$) alone, it is
Hermitian but otherwise arbitrary.

The partition function for this Hamiltonian is derived by implicitly
performing the limit of an infinite number of time slices in its
Suzuki-Trotter representation \cite{pich-etal,rieger-kawa,zamponi}
and yields the imaginary time path integral
\begin{eqnarray}
{\it Z} & = &
{\rm Tr}_{\hat{\sigma},\ooo} \exp(-\beta{\it H})\\
& = &
\int{\it D}\sigma(\tau)\,
\exp(-{\it S}_{\aaa}[\sigma(\tau)])
\label{PI}
\end{eqnarray}
where ${\it S}_{\aaa} [\sigma(\tau)]) = -\ln\,{\rm Tr}_{\ooo}
\exp[\int_0^\beta d\tau\,{\it G}(\sigma(\tau),\ooo,\aaa)]$ and
$\sigma(\tau)$ is now a real valued function of the imaginary time
$\tau\in[0,\beta]$, denoted as a spin-1/2 world line. These world
lines represent realizations of a two-valued Poissonian process that
is sketched in Fig.\ 1a: They are piecewise constant functions
consisting of consecutive segments of spin-up ($\sigma=+1$) and
spin-down ($\sigma=-1$), where the spin-flips occur at stochastic
times $0<\tau_1<\tau_1,\ldots\tau_n$ ($n$ arbitrary) and the interval
lengths $\Delta\tau_i=\tau_{i+1}-\tau_i$ obey a Poissonian statistics
$P(\Delta\tau)=\Gamma^{-1}\exp(-\Gamma\Delta\tau)$ with mean value
$1/\Gamma$ \cite{rieger-kawa}. The path integral (\ref{PI}) can hence
be directly sampled by generating stochastically realizations of such
world lines and accepting them according to their ``Boltzmann''-weight
$\exp(-{\it S}_{\aaa}[\sigma(\tau)])$. More efficient sampling
procedures like cluster algorithms are based on this principle
\cite{rieger-kawa}.

For a general transverse Ising model (without coupling to a
dissipative bath) ${\it G}(\hat{\sigma}^z)$ represents just the
``classical'' energy $E(\hat{\sigma}^z)$ that is diagonal in the
$z$-representation of the spin-1/2 degrees degrees of freedom and
$S[\sigma(\tau)]=\int_0^\beta d\tau\,E(\sigma(\tau))$. This form
holds for an arbitrary number of spins in a transverse field, and for
arbitrary spin-spin interactions.

In the case of the spin-boson model (\ref{SB}) with the spectral
function (\ref{spectral}) the trace over the oscillator degrees of
freedom yields \cite{weiss} ${\it S}_{\aaa}={\it S}_{SB}$ with
\be
{\it S}_{SB}[\sigma(\tau)]=
-\int_0^\beta d\tau \int_0^\tau d\tau'\;
\sigma(\tau)\,{\it K}_\beta(\tau-\tau')\,\sigma(\tau')\;.
\ee
The kernel imposes long-range interactions in imaginary time:
\be
{\it K}_\beta(\tau)=\int_0^\infty d\omega\,\,
\frac{J(\omega)}{\pi}
\frac{\cosh(\frac{\hbar\beta}{2}\omega[1-2\tau/\beta])}
{\sinh(\frac{\hbar\beta}{2}\omega)}\;.
\label{kernel}
\ee
It has the symmetry ${\it K}(\beta-\tau)={\it K}(\tau)$ and the
asymptotics ${\it K}(\tau)\propto\tau^{-(1+s)}$ for $\tau_c\ll\tau\ll\beta$,
where $\tau_c=2\pi/\omega_c$. For $\tau<\tau_c$ the Kernel $K(\tau)$
is regularized via the frequency cut-off $\omega_c$ in
(\ref{spectral}) and approaches a constant for $\tau\to0$.

An efficient way of sampling the path integral is a cluster algorithm
based on \cite{rieger-kawa}. It is generalization of the
Swendsen-Wang cluster algorithm \cite{swendsen-wang} to continuous time
world lines, in which not individual spins but the world line segments
are connected during the cluster-forming procedure, and has to
incorporate the long-range interactions \cite{luijten}. It is sketched
in Fig. 1b: Starting from a world line configuration $\sigma(\tau)$
new potential spin-flip sites are introduced according to a Poissonian
statistics, then all segments are pairwise ``connected'' with
probability
\be
p(s_I,s_{II}) = 1 - \exp\left(-2\,
\int_a^b d\tau \int_c^d d\tau' \, {\it K}_\beta(\tau-\tau')
\right)\;.
\label{prob}
\ee
where $a,b$ and $c,d$ denote the limits of segment $S_I$ and $S_{II}$,
respectively. Finally the connected clusters are identified and flipped
with probability 1/2. All potential spin-flip times that do not
represent real spin-flips are then removed.
\begin{figure}
\includegraphics[width=\linewidth]{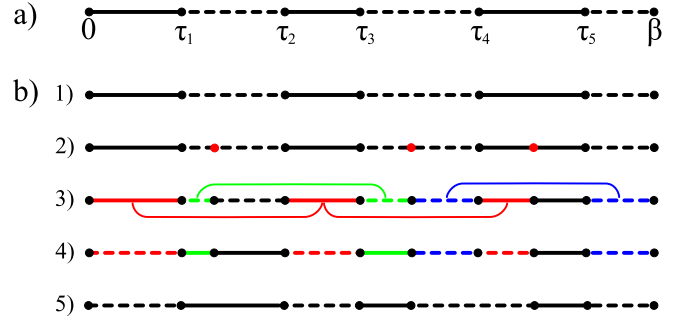}
\caption{(Color online) {\bf a} Realization of an imaginary time
world line of a spin-1/2 in a transverse field. {\bf b)} Sketch of the
continuous time cluster update: 1) Starting configuration. 2) Random
insertion of new potential spin flips (red dots) with Poissonian
statistics 3) Connection of segments with probabilities given by
eq.(\ref{prob}). Different colors indicate the resulting clusters.
4) Each cluster is flipped with probability 1/2 (the blue one was
not flipped). 5) Resulting new imaginary time world line.}

\end{figure}
We implemented this algorithm and tested it by comparing results with
those obtained with conventional Monte-Carlo procedures in discrete
imaginary time extrapolated to an infinite number of time-slices. We
analyzed the sampling characteristics of the algorithm for the kernel
(\ref{kernel}) with (\ref{spectral}) over the whole range $0<s<1$ and
found that on average after 5 updates as sketched in Fig.1b the world
line configuration are statistically independent from the starting
configuration. The data presented below represent averages over
$10^5$-$10^6$ cluster updates.

To study the phase transition in the sub-ohmic spin-boson model
($s<0.5$) we utilize the finite-$\beta$ scaling forms for
thermodynamic observables close to the critical point $\alpha=\alpha_c$
\be
\langle{\it O}\rangle_{T,\alpha} =
\beta^{x_{\it O}} g_{\it O}(\beta^{y_t^*}\delta)\;,
\ee
where $\delta=(\alpha-\alpha_c)/\alpha_c$ denotes the distance from the
critical point, $x_{\it O}$ and $g_{\it O}$ are the scaling exponent
and scaling function of the observable ${\it O}$, respectively. The
exponent $y_t^*$ is $1/\nu$ below the upper critical dimension
($s>1/2$), $\nu$ being the correlation length exponent, and
$y_t^*=1/\nu+(1/2-s)$ above the upper critical dimension ($s<1/2$).
\cite{luijten}.
\begin{figure}
\includegraphics[width=\linewidth]{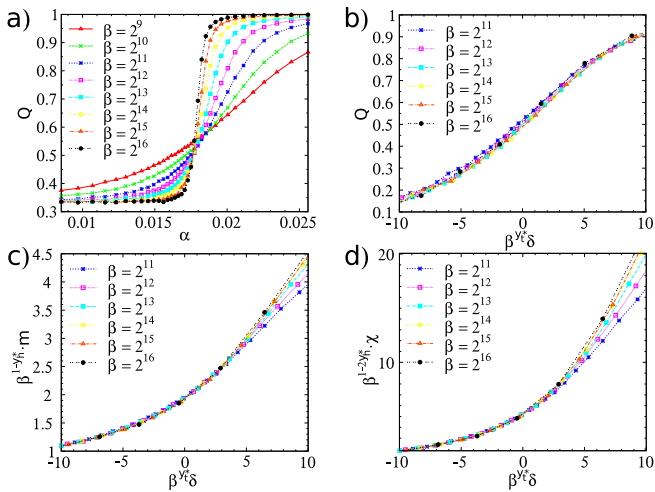}
\caption{Results for the spin boson model for
$s=0.2$ and $\Delta=0.1$. {\bf a)} Moment-ratio $Q$ as a
function of the coupling constant $\alpha$ for different values of
$\beta$. The critical coupling is at $\alpha_c=0.0175\pm0.0002$. {\bf
b-c)} Finite $\beta$-scaling for the moment-ratio $Q$, magnetization
$m$ and susceptibility $\chi$ ($\delta=(\alpha-\alpha_c)/\alpha_c$,
with $\alpha_c$ from a). The values for the critical exponents are
$y_t^*=0.5$, $y_h^*=0.75$. For large positive values of the scaling
variable corrections to scaling are stronger.  }
\end{figure}
We use the dimensionless ratio of moments 
$Q=\langle m^2\rangle^2/\langle m^4\rangle$, which has $x_Q=0$ and is therefore
asymptotically independent of temperature at $\delta=0$, to locate the
critical point $\alpha_c$ as shown for $s=0.2$ in Fig.2a. This estimate
for $\alpha_c$ is then used to perform the finite-$\beta$ scaling
analysis for $Q$, $Q=\tilde{Q}(\beta^{y_t^*}\delta)$ the magnetization
$m=\langle|m|\rangle=\beta^{y_h^*-1}\tilde{m}(\beta^{y_t^*}\delta)$ and
the susceptibility $\chi=\beta\langle m^2\rangle=
\beta^{2y_h^*-1}\tilde{\chi}(\beta^{y_t^*}\delta)$, where $y_h^*$ is the
magnetic exponent. The data collapse that one obtains with the
mean-field values for the exponents $y_t^*$ and $y_h^*$
\be
y_t^*=1/2\,,\quad y_h^*=3/4
\label{exps}
\ee
is good, as shown Fig.2b-d. 
\begin{figure}
\includegraphics[width=\linewidth]{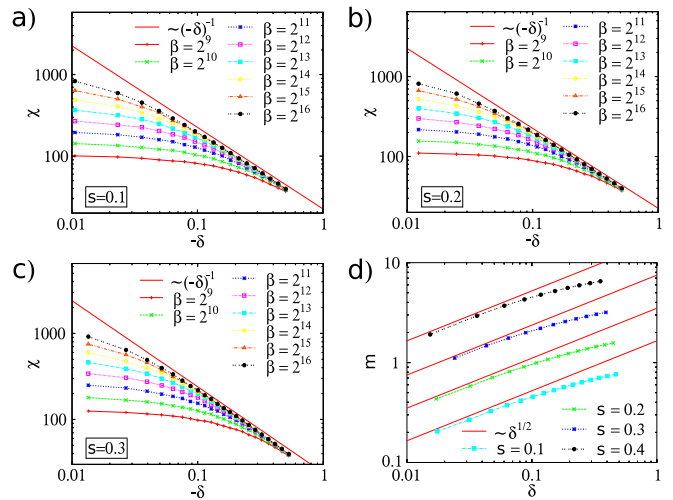}
\caption{{\bf a-c)} Data for the susceptibility $\chi$ as a function
of the distance from the critical point $\delta>0$ (i.e. in the
delocalized phase) for $s=0.1$ (a), $s=0.2$ (b) and $s=0.3$ for
different values of $\beta$ ($\Delta$ and $\omega_c$ as in Fig.2). For
increasing inverse temperature $\beta$ the data points approach the
straight line, which is the zero temperature behavior
$\chi\propto\delta^{-1}$. {\bf d)} Magnetization $m$ as a function of
$\delta>0$ (i.e. in the localized phase) for $\beta=2^{16}$ for
different values of $s$ (multiplied with $2$, $4$ and $8$ for $s=0.2$,
$s=0.3$ and $s=0.4$, respectively, for better visibility) The
straight lines are guides for the eye proportional to the zero
temperature behavior $(\alpha-\alpha_c)^{-1/2}$. Shown are only the
data that are free from finite-$\beta$ corrections. }
\end{figure}

At the critical point $\alpha=\alpha_c$ the scaling forms predict
$\chi\propto T^{-x}$ with $x=2y_h^*-1=1/2$, which is clearly
confirmed by our data displayed in Fig.2d: $\chi\cdot T^{1/2}$
collpaes onto one point at $\delta=0$. Moreover the scaling forms
imply at $T=0$ $\chi\propto|\alpha-\alpha_c|^{-\gamma}$ with
$\gamma=(2y_h^*-1)/y_t^*=1$, which is demonstrated in Fig.3a-c for
different values of $s<1/2$., and
$m\propto(\alpha-\alpha_c)^{\beta_m}$ for $\alpha>\alpha_c$ with
$\beta_m=-(y_h^*-1)/y_t^*=1/2$, which is demonstrated in Fig.3d.

Next we allow for an unbiased fit of the critical exponents to our
data, including corrections to scaling as in \cite{luijten}. We
determined $y_t^*$ and $y_h^*$ by finite-$\beta$ scaling of $\partial
Q/\partial\alpha\,(\alpha=\alpha_c)\propto\beta^{y_t^*}$ and
$\chi(\alpha=\alpha_c)\propto\beta^{2y_h^*-1}$. The results confirm
(\ref{exps}) within the error bars for the whole range of $s<1/2$ that
we studied. Only close to $s=1/2$ the finite-$\beta$ scaling analysis
is impeded by the presence of logarithmic corrections at the upper
critical dimension.  Fig.4a-b shows the resulting estimates for the
exponents $1/\nu=y_t^*-(1/2-s)$ and $\beta_m=-(y_h^*-1)/y_t^*$ as a
function of $s$ in comparison with the NRG predictions of \cite{vojta}. 

\begin{figure}[t]
\includegraphics[width=\linewidth]{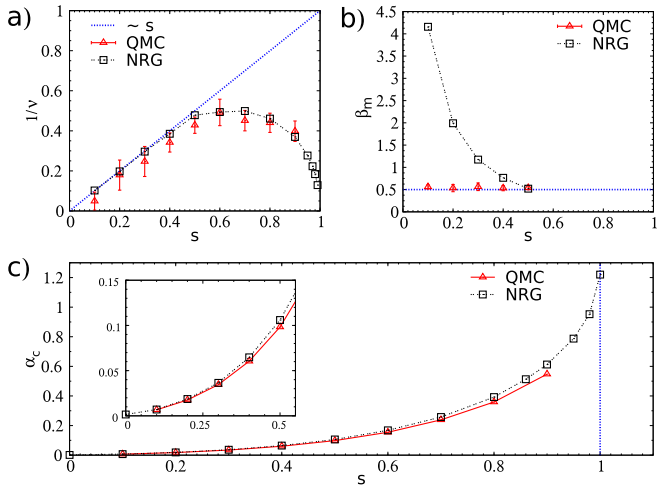}
\caption{
{\bf a-b)} Numerical estimates of the critical exponents $1/\nu$ and
$\beta_m$ as a function of $s$. Triangles: QMC result (obtained from
finite temperature scaling of the QMC data as described in the text);
squares: RG results (from \cite{vojta}, compare also \cite{kevin});
straight lines: mean-filed values for $s<1/2$. {\bf c)} Critical
coupling strength $\alpha_c$ as a function of $s$ diagram for the
spin-boson model (\ref{SB}-\ref{spectral}) with cut-off frequency
$\omega_c=1$ and tunnel matrix element $\Delta=0.1$. Triangles: QMC
result; squares: NRG results for fixed NRG discretization parameter
$\Lambda=2$ (from
\cite{BTV}). Performing the limit $\Lambda\to1$ moves the NRG-estimates 
for $\alpha_c$ slighlty downward.}
\end{figure}

Although our results for the critical exponents of the sub-ohmic bath
obtained with our continuous imaginary time algorithm deviate from the
NRG prediction, results for the phase diagram match: In Fig.4c our
estimates for the critical coupling $\alpha_c$ are compared with those
obtained with the NRG method \cite{vojta}, they agree very well.

We confirmed the scenario described here for other values of $\Delta$
and $\omega_c$, and also for smooth frequency cut-offs as well as for
other kernels (\ref{kernel}), like one that has a regularization in
time (${\it K}(\tau)=0$ for $\tau<\tau_c$) rather than in
frequency. We also found that the limit $\omega_c\to\infty$ (or
$\tau_c\to0$) exists and is approached smoothly and fast, and conclude
that, concerning the critical exponents, the regularization does not
play a significant role.

We also implemented a conventional QMC algorithm in
discrete time (with a finite number of Trotter time slices $M$) and
found that for any fixed value of $\Delta\tau=\beta/M$ mean-field
exponents describe perfectly the scaling at the critical point for
$s<1/2$ (see also \cite{luijten,remark}). Moreover we found that the
extrapolation $M\to\infty$ of numerical data for $Q$, $m$ and $\chi$
obtained for fixed $M$ reproduces perfectly the results obtained with
our continuous imaginary time cluster algorithm and that the
convergence is smooth and fast (with $1/M$, as expected).

Our conclusion therefore is that the quantum-to-classical mapping does
not fail in the sub-ohmic spin-Boson model. The question remains, why
the NRG calculation presented in \cite{BTV,vojta} yields apparently
correct results for quantities like the critical coupling, i.e. the
phase diagram (see Fig. 4c), but fails to predict the correct critical
exponents in the case $s<1/2$.

We believe the problem is rooted in a shortcoming of the present NRG
implementation.  As detailed in Ref.~\onlinecite{BLTV}, due to the
truncation of the bosonic Hilbert space, the NRG -- while correctly
describing the delocalized phase and the critical point -- it is
unable to capture the physics of localized phase of the spin-boson
model for $s<1$.  Technically, a finite expectation value
$\langle\sigma_z\rangle$ is accompanied by a mean shift of the bath
oscillators which diverges in the low-energy limit.  Hence, the NRG
results are expected to be reliable as far as they do not involve
properties of the localized fixed point.

The analysis of critical exponents in Ref.~\onlinecite{vojta} now
assumed that all exponents are properties of the critical fixed
point.  However, this assumption is {\em invalid} for the
order-parameter related exponents $\beta_m$ and $\delta_m$ {\em if} the
critical fixed point is Gaussian (like in a $\phi^4$ theory above its
upper critical dimension).  Then, the order parameter amplitude is
controlled by a dangerously irrelevant variable, and $\beta_m$ and
$\delta_m$ are properties of the flow towards the localized fixed point,
which in turn is not correctly captured by NRG.
(Note that $\delta_m$ involves the non-linear field response at
criticality, which is undefined for a purely Gaussian theory.)
Therefore, the values of $\beta_m$ and $\delta_m$ extracted from (present)
NRG calculations are unreliable.

Considering that the NRG calculations nevertheless gave well-defined
power laws which were moreover consistent with hyperscaling, it is
worth asking for the underlying reason.  We conjecture that the
artificial Hilbert-space truncation, which determines the flow to the
``wrong'' localized fixed point and limits both the field response and
the condensate amplitude, is equivalent to an operator which is
exactly marginal at criticality in the $\phi^4$ language.  Near
criticality, this has no consequences below the upper critical
dimension, $s>1/2$, as the quartic interaction is relevant here, but
for $s<1/2$ the marginal operator instead dominates over the quartic
term.  It is easy to show that an exactly marginal coupling leads to
$y_h^*=(1+s)/2-y_i$ with $y_i=0$ such that hyperscaling is fulfilled,
while $y_t^*$ takes its mean-field value -- this is what characterizes
the set of NRG critical exponents \cite{vojta}.
(The correct result $y_h^*=3/4$ for $s<1/2$ implies $y_i=(2s-1)/4$
arising from the dangerously irrelevant variable.)
The above reasoning is supported by analyzing
fermionic impurity models, which naturally have the property that a
Hilbert space constraint limits the field response.  For instance, a
resonant-level model with power-law bath density of states, which is
controlled by a stable intermediate-coupling fixed point, shows
hyperscaling for all bath exponents \cite{LFMV}.

Finally, the analytical RG argument in Ref.~\onlinecite{vojta},
based on an epsilon-expansion for small $s$, predicted non-mean-field
exponents obeying hyperscaling for a related reason:
While the RG equations (9)-(11) of Ref.~\onlinecite{vojta} are correct,
the subsequent analysis overlooked the presence of the dangerously irrelevant
variable, resulting again in the incorrect $y_h^*=(1+s)/2$.

To conclude we have, with the help of an efficient and accurate
continuous time cluster Monte-Carlo algorithm, shown that the
quantum-to-classical mapping {\it is} valid in the case of the
sub-ohmic spin-boson model. The presence of a dangerously irrelevant
variable for $s<1/2$ impedes the correct extraction of the critical
exponents with current versions of the NRG method - work on its
extension to produce reliably the necessary determination of magnetic
observables in the localized phase is in progress.

\end{document}